# The Renormalized Trajectory of the O(N) Non-linear Sigma Model

Wolfgang Bock[a] [*] and Julius Kuti[a]

[a]University of California, San Diego, Dept. of Physics, 9500 Gilman Dr., La Jolla, CA 92093-0319, USA

The renormalized trajectory (RT) is determined from two different Monte Carlo renormalization group techniques with $\delta$-function block spin transformation in the multi-dimensional coupling parameter space of the two-dimensional non-linear sigma model with O(3) symmetry. At a correlation length $\xi \approx$ 3-5, the RT is shown to break away from the straight line of the fixed point trajectory (FPT) which is orthogonal to the critical surface and originates from the ultraviolet fixed point (UVFP). The large $N$ calculation of the RT is also presented in the coupling parameter space of the most general bilinear Hamiltonian. The RT in the large $N$ approximation exhibits a similar shape with the sharp break occurring at a somewhat smaller correlation length.

## 1. INTRODUCTION

A sketch of the RT in the phase diagram of asymptotically free (AF) field theories is shown in Fig. 1 as plotted in the infinite dimensional space of couplings $K_i$. The inverse coupling $1/K_1$ of the standard lattice action is singled out to label the horizontal axis. The critical manifold is given in Fig. 1 by the $K_1 = \infty$ plane and UVFP designates the AF fixed point on the critical surface. The continuum theory at finite lattice correlation length is defined by the RT which flows along the unstable direction from the UVFP to the high temperature fixed point (HTFP) with vanishing correlation length at $K_i = 0$ [1].

It has been known for a long time that, in contrast with the standard lattice action, each point on the RT defines a perfect lattice action [2] which is free of cutoff effects at any finite correlation length [1]. The determination of this perfect lattice action, however, has not been considered feasible in the past, mainly because a realistic approximation to the RT in the infinite dimensional coupling parameter space was thought to require the uncontrolled truncation of a large number of terms in the blocked Hamiltonian.

Considerable progress has been made recently by Hasenfratz and Niedermayer [2] who realized that the fixed point lattice action of the UVFP can be determined from a classical saddle point problem in AF field theories. The FPT which originates from the UVFP is suggested to be a good approximation to the RT at sufficiently large correlation lengths [2] (dashed line at $K_1 < \infty$ in Fig. 1.) It was also demonstrated that the fixed point action (FPA) can be rendered very short-ranged by optimizing the block-spin renormalization group transformation with the expectation that an approximate FPT which is based on a truncated FPA, with the very small long range couplings neglected, will exhibit almost no cutoff dependence at large correlation lengths.

It remains an important question to determine the crossover region in the correlation length where the RT departs from the FPT. The performance of the approximate FPT has first been tested by Hasenfratz and Niedermayer in the 2d non-linear O(3) sigma model which is known to be AF [2]. They carried out a pilot study using a truncated FPT with 24 different couplings. Several tests showed that the residual cutoff effects were not visible even down to a correlation length of five suggesting that truncation effects are negligible. This indirectly implies that the FPT runs close to the RT in an extended range of the lattice correlation length.

The RT eventually has to break away from the FPT since it flows into the HTFP as depicted in Fig. 1. For practical applications, it is important to determine the crossover region in the correlation length where the RT breaks away from the FPT. In this contribution we determine numerically the position of the RT in a finite dimensional

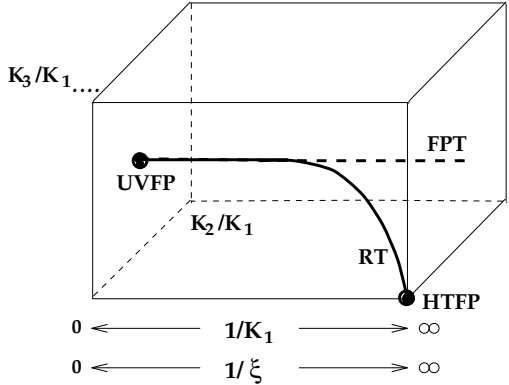

Figure 1. *Phase digram of AF field theories.*

subspace of the infinite dimensional coupling parameter space and compare it with the FPT.

## 2. THE FIXED POINT ACTION

The path integral of the O(N) non-linear sigma model on the lattice is given by $Z = \int \prod_x \mathcal{D}\phi_x \exp(-\beta\mathcal{H})$, where $x$ is a two-dimensional lattice vector to label the sites, $\phi_x$ is an $N$ component scalar field, and $\int \mathcal{D}\phi_x = \int_{-\infty}^{+\infty} \prod_x d\phi_x \delta(1-\phi_x^2)$ is the O(N) invariant measure. The lattice hamiltonian $\mathcal{H}$ is not unique. The only constraints come from the symmetries of the system and the fact that it ought to reduce in the classical continuum limit to the continuum hamiltonian. The lattice hamiltonian with these constraints can be parametrized as

$$\mathcal{H} = \mathcal{H}_2 + \mathcal{H}_4 + \mathcal{H}_6 + \dots , \quad (1)$$

$$\mathcal{H}_2 = -\frac{1}{2}\sum_r\sum_x \rho(r) \ (1 - \phi_x\phi_{x+r}) , \quad (2)$$

$$\mathcal{H}_4 = \sum_{x_1,x_2,x_3,x_4} c_{x_1,x_2,x_3,x_4} \ (1 - \phi_{x_1}\phi_{x_2})(1 - \phi_{x_3}\phi_{x_4}), (3)$$

where $\mathcal{H}_2$ includes all terms that are bilinear and $\mathcal{H}_4, \dots$ designate all the other terms which are quartic or of higher order in the field variables. The summations in Eqs. (2) and (3) are over lattice vectors $r$, $x$, $x_1$, $x_2$, $x_3$, and $x_4$. The standard lattice action is given by the first two terms in Eq. (2) with $r = (1, 0)$, $(0, 1)$, and $K_1 = \beta\rho(r_1)$.

Let us now consider the following $\delta$-function

Table 1
*The six largest couplings $\rho^*(r)$.*

| $r$ | $\rho^*(r)$ |
|---|---|
| $r_1 = (1, 0)$ | $-3.42839$ |
| $r_2 = (2, 0)$ | $+0.74981$ |
| $r_3 = (1, 1)$ | $+0.32486$ |
| $r_4 = (3, 0)$ | $-0.16871$ |
| $r_5 = (4, 0)$ | $+0.03877$ |
| $r_6 = (3, 1)$ | $-0.01976$ |

block spin transformation,

$$e^{-\beta'\mathcal{H}'(\phi';\rho',c',\dots)} = \int \mathcal{D}\phi P(\phi,\phi')e^{-\beta\mathcal{H}(\phi;\rho,c,\dots)}, \quad (4)$$

$$P(\phi,\phi') = \prod_{x'} \delta\left(\phi'_{x'} - \frac{\sum_{x\in x'}\phi_x}{\|\sum_{x\in x'}\phi_x\|}\right), \quad (5)$$

where a blocked lattice site $x'$ is assigned to a $2 \times 2$ cell of sites $x$ on the unblocked lattice. For details on how the FPA can be computed in AF field theories we refer the reader to Ref. [2]. The authors of Ref. [2] showed that the fixed point couplings $\rho^*(r)$ agree with the ones obtained earlier for the non-interacting model with Gaussian block spin transformation [3],

$$\rho^*(r) = \int_{-\pi}^{+\pi} \frac{d^2p}{(2\pi)^2} \ \rho^*(p) \ e^{-ipr} , \quad (6)$$

$$\rho^*(p) = \left[ \sum_{n_1,n_2=-\infty}^{+\infty} \frac{1}{(p_1+2\pi n_1)^2 + (p_2+2\pi n_2)^2} \right.$$
$$\left. \times \prod_{i=1}^{2} \frac{\sin^2(p_i/2)}{(p_i/2 + n_i\pi)^2} \right]^{-1} , \quad (7)$$

which holds independently of $N$. We have determined the couplings for several lattice vectors $r$ by evaluating Eqs. (6) and (7) numerically. The values of the six largest couplings are given in Table 1. The interactions are indeed very shortranged since the couplings decrease rapidly when the distance $|r|$ between two spins grows.

## 3. LARGE N CALCULATION

In the large $N$ limit, Hirsch and Shenker derived a recursion relation for the blocked two-point function in momentum space [4],

$$G'(p') = \frac{\sum_l G(\frac{p'}{2} + \pi l)\, s^2(\frac{p'}{2} + \pi l)}{\frac{4}{L^2} \sum_{p',l} G(\frac{p'}{2} + \pi l)\, s^2(\frac{p'}{2} + \pi l)}, \quad (8)$$

$$s^2(p') = \prod_{\mu=1}^{2} \frac{\sin^2 p'_\mu}{\sin^2(p'_\mu/2)}, \quad (9)$$

where $l$ is a vector with components equal to 0 or 1. $L^2$ designates the number of points on the lattice before blocking, and $G(p) = [\beta\rho(p) + \lambda]^{-1}$ is the saddle point solution with the mass gap $\sqrt{\lambda}$ determined from the gap equation, $\frac{1}{L^2} \sum_p G(p) = 1$. Eq. (8) can now be iterated to determine the blocked propagators after repeated block-spin transformations. In the subspace of bilinear Hamiltonians we can determine the couplings $\beta\rho(r)$, $\beta'\rho'(r)$, etc. from the inverse propagators by Fourier transformation $\beta\rho(r) = \frac{1}{L^2} \sum_p G(p)^{-1}$, $\beta'\rho'(r) = \frac{4}{L^2} \sum_{p'} G'(p)^{-1}$, etc.

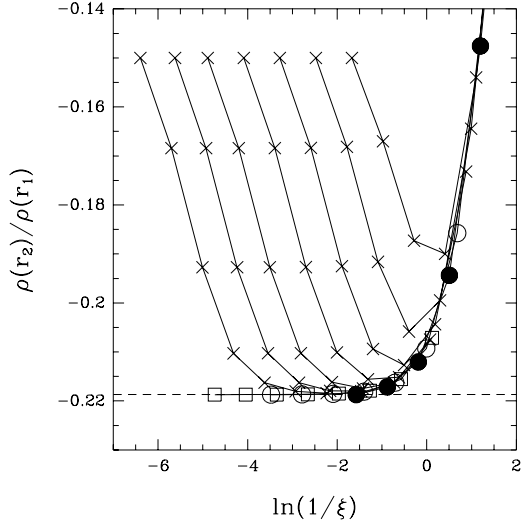

Figure 2. *Flow lines of $\rho(r_2)/\rho(r_1)$ in the large N limit.*

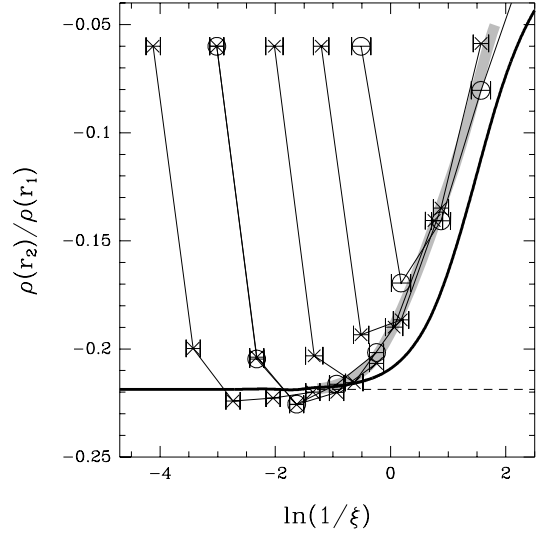

Figure 3. *Flow lines of $\rho(r_2)/\rho(r_1)$ for $N = 3$.*

We have iterated Eq. (8) numerically, using two different expressions for the unblocked propagator. The first choice originates from the bilinear Hamiltonian that includes the four largest couplings in Table 1 and the second choice is the fixed point propagator resulting from Eq. (7). In the first case we start the flow lines, which after several blocking steps approach the RT, at different positions in the four-dimensional coupling parameter space. In the second case, the flow lines start on the FPT and it is interesting to see how the blocked couplings break away from the FPT after a few block spin transformations.

As an example, we have displayed in Fig. 2 the flow lines of the ratio $\rho(r_2)/\rho(r_1)$ as a function of $\ln(1/\xi)$, with $\xi$ designating the correlation length. The plots for the other couplings look very similar and we refer the reader for more details to Ref. [5]. The UVFP is located in this plot at $\ln(1/\xi) = -\infty$ and the HTFP at $\ln(1/\xi) = +\infty$. The ratio $\rho^*(r_2)/\rho^*(r_1)$ has been represented in this graph by the horizontal dashed line. The flow lines which start from the four-coupling Hamiltonian are represented by crosses. Fig. 2 shows that those flow lines that start at large correlation lengths approach the FPT within a few block spin steps indicating that the FPT is indeed a very good approximation to the RT at large correlation lengths. However, a different behavior is seen at small correlation lengths where the flow lines approach the RT which runs into the HTFP with the ratio $\rho(r_2)/\rho(r_1)$ vanishing, as can be shown by the high temperature expansion [5]. Fig. 2 shows that the sharp break from the FPT occurs at a correlation length of $2-3$. The three flow lines that start on the FPT are represented in the graph by squares, full, and open circles.

## 4. NUMERICAL SIMULATION OF THE O(3) MODEL

The renormalization group flows in the previous section were restricted to the subspace of bilinear Hamiltonians in the large $N$ limit. In contrast, the Monte Carlo renormalization group method will allow us to study the exact flows of selected couplings in the O(3) model. Although interaction terms which do not fit on the finite lattice are truncated, their effects are expected to be negligible. The only practical limitation on the method is the signal to noise ratio in measuring very small blocked couplings.

The 1-cluster algorithm has been used to generate unblocked spin configurations on a $256 \times 256$ lattice using a Hamiltonian with the first four couplings of Table 1. The spin configurations have been blocked four times down to a lattice of size $16 \times 16$ using the block spin transformation of Eq.(4). Microcanonical [6] and canonical [7] demon methods have been implemented to infer the blocked couplings from the blocked spin configurations. For more details and a comparison of the two methods, we refer the reader to ref. [5].

To illustrate our results, we have plotted again in Fig. 3 the renormalization group flow of the ratio $\rho(r_2)/\rho(r_1)$ as a function of $\ln(1/\xi)$. Most of the data have been obtained with the microcanonical method (crosses). We have also included two renormalization group flow lines which have been generated by the canonical demon method (circles), one of which with the same initial couplings as a selected microcanonical flow line. Fig. 3 shows that the two flow lines coincide and the flow pattern is very similar to the one seen in Fig. 2. The only difference is that the sharp break from the FPT occurs now at a somewhat larger correlation length. The RT of the O(3) model is traced by the grey line. For comparison, we have also included in the graph the RT in the large $N$ limit (heavy line) (cf. Fig. 2).

## 5. CONCLUSIONS

We have demonstrated in this paper by explicitly computing the RT in the O(3) model that the FPT provides a good approximation to the RT in the region of the coupling parameter space where the correlation length is large. A significant break occurs, however, at a correlation length of about 3-5 where the RT sharply departs from the FPT and flows into the HTFP. The RT in the large $N$ limit exhibits a similar behavior with the sharp break shifted to somewhat smaller correlation length. Although we have only shown the projection of the exact RT in the $K_1$-$K_2$ plane, several other projections of the flows were determined as well [5].

Both ends of the RT can be determined without the utilization of the Monte Carlo renormalization group. The RT in the large correlation length regime, near the UVFP, is well approximated by the FPT, while the RT in the regime of small correlation lengths, near the HTFP, can be obtained from a high temperature expansion. It is likely that both approximations will break down in the crossover region where the Monte Carlo renormalization group technique may remain the only useful tool. It is not clear whether this regime will be important in practical applications.

The studies we presented will be useful to extend to non-Abelian lattice gauge theories in four dimensions [8].

This work was supported by the DOE under grant DE-FG03-91ER40546.